\documentclass[acmlarge,timestamp,screen]{acmart}
\pdfoutput=1

\title{Kanren Light}
\subtitle{A Dynamically Semi-Certified Interactive Logic Programming System}
\author{Marco Maggesi}
\email{marco.maggesi@unifi.it}
\author{Massimo Nocentini}
\email{massimo.nocentini@unifi.it}
\affiliation{%
\institution{University of Florence}
\department{Departement of Mathematics and Computer Science}
\streetaddress{Viale Morgagni, 65}
\city{Florence}
\postcode{50134}
\country{Italy}}
\setcopyright{acmcopyright}
\copyrightyear{2020}
\acmYear{2020}

\acmConference[miniKanren 2020]{25th ACM SIGPLAN International Conference on Functional Programming}{August 23--28, 2020}{Jersey City, New Jersey}

\keywords{miniKanren, HOL Light, Certified, Theorem Proving, Program verification}

  \ccsdesc[300]{Theory of computation~Interactive computation}
  \ccsdesc[300]{Theory of computation~Streaming models}
  \ccsdesc[500]{Theory of computation~Logic and verification}
  \ccsdesc[500]{Theory of computation~Automated reasoning}

\begin{document}

\begin{abstract}
We present an experimental system strongly inspired by miniKanren,
implemented on top of the tactics mechanism of the HOL~Light
theorem prover. Our tool is at the same time a mechanism for enabling the logic
programming style for reasoning and computing in a theorem prover,
and a framework for writing logic programs that produce solutions
endowed with a formal proof of correctness.
\end{abstract}

\maketitle

\section{Introduction}
\label{sec:introduction}

In straightforward terms, the computation of a logic program evolves by
refining a substitution seeking for solutions of a unification problem.  This
has been made explicit in the Kanren approach
\cite{10.5555/1834996,Hemann:2013,friedman2018reasoned}, where programs are
described by composing (higher-order) operators that act on streams of
substitutions. Such a methodology allows for a streamlined approach to logic
programming; however, the intended semantics and the correctness of a Kanren
program rest entirely on the meta-theoretic level.

We propose a framework that extends the Kanren approach to a system that
computes both candidate substitutions and corresponding certificates of
correctness with respect to a given specification. Such certificates will be
formally verified logical truths synthesized using a theorem prover.

Our setup is based on the HOL~Light theorem prover
\cite{10.5555/646184.682934}, in which we extend the currently available tactic
mechanism with three basic features: (i)~the explicit use of meta-variables,
(ii)~the ability to backtrack during the proof search, (iii)~a layer of tools
and facilities for interfacing with the underlying proof mechanism.

The basic building block of our framework are ML procedures that we call
\emph{solvers}, which are a generalization of HOL tactics and are~--as well as
tactics-- meant to be used compositionally in order to define arbitrarily
complex proof search strategies.

We say that our approach is \emph{semi-certified} because
\begin{itemize}
\item on the one hand, the synthesized solutions are formally proved
  theorems, hence their validity is guaranteed by construction;
\item on the other hand, the completeness of the search procedure
  cannot be enforced in our framework and consequently has to be
  ensured by a meta-reasoning.
\end{itemize}
Moreover, we say that our system is \emph{dynamically} semi-certified,
because the proof certificate is built at run-time.

At the present stage, our implementation is intended to be a testbed
for experiments and further investigation on this reasoning paradigm.
Section~\ref{sec:code} gives some further information on our code.

\section{A word about the HOL~Light theorem prover}
\label{sec:hol-light}

In the HOL system, there are two fundamental datatypes called \emph{term} and
\emph{theorem}.  Terms model fragments of (well-formed) mathematical
expressions.  Theorems are Boolean terms that are proved correct according to
a fixed set of logical rules.  Examples of both a term and a theorem in the
concrete syntax of HOL~Light are
\[
\verb|`2 + 2`|\qquad \text{and}\qquad \verb.|- 2 + 2 = 4..
\]
Notice that terms are written enclosed in backquotes while theorems use the
entailment symbol \verb+|-+\,.

A theorem \verb+as |- b+ is a first-class value composed by a list of terms
\verb|as| and a body \verb|b|, for instance
\begin{verbatim}
# ARITH_SUC;;
val it : thm =
  |- (!n. SUC (NUMERAL n) = NUMERAL (SUC n)) /\
     SUC _0 = BIT1 _0 /\
     (!n. SUC (BIT0 n) = BIT1 n) /\
     (!n. SUC (BIT1 n) = BIT0 (SUC n))
\end{verbatim}
is synthetized via regular function call of \verb|prove| that actually consumes
the theorem's body and the corresponding proof,
\begin{verbatim}
let ARITH_SUC = prove
 (`(!n. SUC(NUMERAL n) = NUMERAL(SUC n)) /\
   (SUC _0 = BIT1 _0) /\
   (!n. SUC (BIT0 n) = BIT1 n) /\
   (!n. SUC (BIT1 n) = BIT0 (SUC n))`,
  REWRITE_TAC[NUMERAL; BIT0; BIT1; DENUMERAL ADD_CLAUSES]);;
\end{verbatim}
The Boolean connectives `$\wedge$', `$\vee$', `$\Longrightarrow$' are
represented in ASCII encoding \verb+/\+, \verb+\/+ and \verb+==>+, respectively.
Universal and existential quantifier $\forall x.\, P x$ and
$\exists x.\, P x$ are denoted with exclamation and interrogation marks:
\verb+!x. P x+ and \verb+?x. P x+.  Other syntactic elements are borrowed
from the ML world, such us the notation for concrete lists
\verb+[x1;+\dots \verb+]+.

As the name suggests, HOL~(\emph{Higher-Order Logic})
implements a higher-order language based on a variant of the typed lambda calculus.
Hence, in a rough comparison with classical logic programming languages,
our system is closer to $\lambda$Prolog~\cite{Miller91alogic} than the usual (first-order)~Prolog.

Interactive proofs in HOL~Light are performed by running \emph{tactics}
that operate on a context called \emph{goal}, which represents the
intermediate status of the current logical reasoning.  There are simple tactics that
model basic logical inference steps as well as sophisticated tactics that
implement powerful decision procedures.

From the theorem proving perspective, our work consists of extending the
HOL~Light's tactic mechanism by introducing specific ideas coming
from the miniKanren methodology.  The resulting system allows the user
to build proof scripts either with tactics or solvers, and the resulting
theorems will be available to the programming environment regardless
of which proof mechanism has been utilized.  In particular, the new
theorems can be built on top of the standard library of HOL Light,
populated by several thousand theorems.

\section{A simple example}
\label{sec:a-simple-example}

To give the flavor of our framework, we show an example of how to perform
simple computations on lists.  Let us consider the problem of computing the
concatenation of two lists \verb|[1; 2]| and \verb|[3]|.  One idiomatic way to
approach this problem in HOL is by using \emph{conversions}
\cite{PAULSON1983119}.  Conversions are ML procedures that receive as input a
term $t$ and output a theorem of the form \verb+|- +$t$\verb+ = +$t'$.  The
term $t'$ is the \emph{result} of the computation, and the theorem itself is
the \emph{certificate} that guarantees its correctness.  Let us show first how
conversions are used before describing how one can perform the same task using
our framework.

In HOL~Light, one has the constant \verb+APPEND+ and the
equational theorem (of the same name) that characterize it
\begin{verbatim}
  |- (!l. APPEND [] l = l) /\
     (!h t l. APPEND (h :: t) l = h :: APPEND t l)
\end{verbatim}
We can then use the conversion \verb+REWRITE_CONV+ which performs the
rewriting.  The ML command is
\begin{verbatim}
  # REWRITE_CONV [APPEND] `APPEND [1;2] [3]`;;
\end{verbatim}
which produces the theorem
\begin{verbatim}
  |- APPEND [1; 2] [3] = [1; 2; 3]
\end{verbatim}

Our implementation allows us to address the same problem from a logical point
of view.  We start by recalling two theorems that are proved -- via list
structural induction -- during the bootstrap procedure of HOL Light, namely
\begin{verbatim}
# APPEND_NIL;;
val it : thm = |- !l. APPEND [] l = l
\end{verbatim}
and
\begin{verbatim}
# APPEND_CONS;;
val it : thm =
  |- !x xs ys zs. APPEND xs ys = zs
                  ==> APPEND (x :: xs) ys = x :: zs
\end{verbatim}
to give the logical rules, in form of Horn clauses, that characterize the \verb|APPEND|
operator.  Then we define a \emph{solver}
\begin{verbatim}
let APPEND_SLV : solver =
  REPEAT_SLV (CONCAT_SLV (ACCEPT_SLV APPEND_NIL)
                         (RULE_SLV APPEND_CONS));;
\end{verbatim}
which implements the most obvious strategy for proving a relation of the form
\verb|`APPEND x y = z`| by structural analysis on the list \verb|`x`|.  The
precise meaning of the above code will be clear later; however, this can be
seen as the direct translation of the Prolog program
\begin{verbatim}
append([],X,X).
append([X|Xs],Ys,[X|Zs]) :- append(Xs,Ys,Zs).
\end{verbatim}

Then, the problem of concatenating the two lists is described by the
term
\begin{verbatim}
`??x. APPEND [1;2] [3] = x`
\end{verbatim}
where the binder \verb|`(??)`| is a syntactic variant of the usual
existential quantifier \verb|`(?)`|, which introduces the
\emph{meta-variables} of the \emph{query}.

The following command
\begin{verbatim}
list_of_stream
  (solve APPEND_SLV
         `??x. APPEND [1; 2] [3] = x`);;
\end{verbatim}
runs the search process where (i)~the \verb|solve| function starts the proof
search and produces a stream (i.e., a lazy list) of \emph{solutions} and
(ii)~the outermost \verb|list_of_stream| transforms the stream into a list.

The output of the previous command is a single solution which is
represented by a pair where the first element is the instantiation for
the meta-variable \verb|`x`|and the second element is a HOL theorem
\begin{verbatim}
val it : (term list * thm) list =
  [([`x = [1; 2; 3]`], |- APPEND [1; 2] [3] = [1; 2; 3])]
\end{verbatim}
Since the theorem is the instantiation of the original query term,
it certifies the correctness of the solution.

Now comes the interesting part: as in logic programs, our search
strategy (i.e., the \verb|APPEND_SLV| solver) can be used for backward
reasoning. Consider the variation of the above problem where we want to enumerate
all possible splits of the list \verb|[1; 2; 3]|.  This can be done by
simply changing the goal term in the previous query:
\begin{verbatim}
# list_of_stream
    (solve APPEND_SLV
           `??x y. APPEND x y = [1;2;3]`);;

val it : (term list * thm) list =
  [([`x = []`; `y = [1; 2; 3]`],
    |- APPEND [] [1; 2; 3] = [1; 2; 3]);
   ([`x = [1]`; `y = [2; 3]`],
    |- APPEND [1] [2; 3] = [1; 2; 3]);
   ([`x = [1; 2]`; `y = [3]`],
    |- APPEND [1; 2] [3] = [1; 2; 3]);
   ([`x = [1; 2; 3]`; `y = []`],
    |- APPEND [1; 2; 3] [] = [1; 2; 3])]
\end{verbatim}
The system finds the above solutions by filtering and refining a stream
of substitutions, precisely in the same way it is done in
any typical miniKanren implementation; eventually, the interesting part is the
associated theorems that are synthesized.

\section{A library of solvers}
\label{sec:library-solvers}

Our framework is based on ML procedures called \emph{solvers} which
generalize classical HOL tactics in two ways: (i)~they facilitate the
manipulation of meta-variables (and their associated substitutions) in the
goal\footnote{The tactic mechanism currently implemented in HOL~Light already
provides basic support for meta-variables in goals.  However, it seems to be
used only internally in the implementation of the intuitionistic tautology
prover \texttt{ITAUT\_TAC}.} and (ii)~they allow the proof search to backtrack.
Before digging into the description of what a solver is, we warn the reader
that the word \emph{goal} has a different meaning in miniKanren and HOL. For the
former, a goal is a function that consumes a substitution and produces a stream
of substitutions; for the latter, a goal is a pair of (already proved)
assumptions and a term that still has to be proved. From now on, we will use
the word \emph{goal} in the sense of HOL.

For the sake of completeness, it is worth to describe the differences among
goals, tactics, and solvers.

On the one hand, the refinements that a miniKanren goal does on substitutions are
performed by a HOL tactic which takes a HOL goal apart into a tuple
$(\mathcal{M}, \mathcal{S}, f)$, where $\mathcal{M}$ is a set collecting the
introduced meta-variables so far, $\mathcal{S}$ is a list of (sub)goals, and $f$
is a function that certifies the performed refinement. The usual HOL routine is
to push and pop those tuples in a stack that represents the steps left to prove
the claimed term -- whenever the stack gets empty, the proof is completed.

On the other hand, a \textit{solver} is a function that consumes a HOL goal as
well as a tactic does, and produces a \textit{stream} of such tuples that
actually allows us to equip HOL~Light with backtracking.
To tie the knot, solvers
extend tactics in the sense that every HOL tactic can be ``promoted'' into a
solver using the ML function
\begin{verbatim}
TACTIC_SLV : tactic -> solver
\end{verbatim}

We provide a library of basic solvers, usually having a name that ends in
\verb|_SLV|. For the rest of the paper, the following elementary solvers
\begin{itemize}
	\item \verb|RULE_SLV : thm -> solver|, that implements the backward chaining rule;
	\item \verb|ACCEPT_SLV : thm -> solver|, that solves a goal by unifying with the supplied theorem;
	\item \verb|CONJ_SLV : solver|, that splits a goal using the introduction rule of the conjunction;
	\item \verb|REFL_SLV : solver|, that solves a goal which is an equation by unifying of the left- and right-hand sides;
	\item \verb|ALL_SLV : solver|, that leaves the goal unmodified.
\end{itemize}

Please note that, as in miniKanren systems, the unification procedure employed is
not hard-wired by our framework, and each solver can implement its own unification
strategy.  We see two main interesting variants that one would have at
disposal.  The first one is to use pattern matching instead of unification;
this would allow for a mechanism of input/output modes as in certain Prolog
implementations.  The second one would be to use a higher-order unification
algorithm to unleash the full expressivity of the underlying higher-order
language.




\emph{Solvers are highly compositional}, as tactics in HOL and goals in miniKanren
are, and complex solvers can be built from simpler ones using high-order
functions.  For instance, given two solvers $s_1$ and $s_2$ the solver
combinator \verb|CONCAT_SLV| make a new solver that collect sequentially all
solutions of $s_1$ followed by all solutions of $s_2$.  This is the most basic
construction for introducing backtracking into the proof strategy. The solver
\verb|COLLECT_SLV| iterates \verb|CONCAT_SLV| over a list of solvers.  Two
other high-order solvers are (i)~\verb|THEN_SLV : solver -> solver -> solver|
which combines sequentially two solvers and (ii)~\verb|REPEAT_SLV : solver -> solver|
that keeps applying a given solver.  Unlike Prolog, miniKanren uses a
complete search strategy by default and that is provided in our system as well
by the solver
\begin{verbatim}
let INTERLEAVE_SLV (slvl:solver list) : solver =
  if slvl = [] then NO_SLV else
  mergef_stream slvl [];;
\end{verbatim}
that relies on the stream combinator
\begin{verbatim}
mergef_stream : ('b -> 'a stream) list -> (unit -> 'a stream) list -> 'b -> 'a stream
\end{verbatim}
which merges two lists of streams by interleaving each one of them.

\emph{Solvers (as for classical HOL tactics) can be used interactively} by means of
the following essential commands:
\begin{itemize}
  \item \verb|gg |$\langle \mathit{term}\rangle$ starts a new goal;
  \item \verb|ee |$\langle \mathit{solver}\rangle$ applies a solver to the current goal state;
  \item \verb|bb ()| restores the previous goal state (i.e., undo the previous \verb|ee| command);
  \item \verb|top_thms ()| returns the stream of solutions found.
\end{itemize}
Here is an example of interaction.
We first introduce the goal, notice the use of the binder \verb|(??)| for the
meta-variable \verb|x|:
\begin{verbatim}
# gg `??x. 2 + 2 = x`;;
val it : mgoalstack =
`2 + 2 = x`
Metavariables: `x`,
\end{verbatim}
one possible solution is by using reflexivity that closes the proof
\begin{verbatim}
# ee REFL_SLV;;
val it : mgoalstack = No sub(m)goals
\end{verbatim}
and allows us to form the resulting theorem
\begin{verbatim}
# list_of_stream(top_thms());;
val it : (instantiation * thm) option list =
  [Some (([], [(`2 + 2`, `x`)], []), |- 2 + 2 = 2 + 2)]
\end{verbatim}

Now, if one want to find a different solution, we can restore the
initial state
\begin{verbatim}
# bb();;
val it : mgoalstack =
`2 + 2 = x`
Metavariables: `x`,
\end{verbatim}
then use a different solver, for instance by unifying with the
equational theorem \verb?|- 2 + 2 = 4?, which can be automatically
proved using the HOL procedure \verb|ARITH_RULE|,
\begin{verbatim}
# ee (ACCEPT_SLV(ARITH_RULE `2 + 2 = 4`));;
val it : mgoalstack = No sub(m)goals
\end{verbatim}
and, again, take the resulting theorem
\begin{verbatim}
# list_of_stream(top_thms());;
val it : (instantiation * thm) option list =
  [Some (([], [(`4`, `x`)], []), |- 2 + 2 = 4)]
\end{verbatim}

Finally, we can change the proof strategy to find both solutions by
using backtracking
\begin{verbatim}
# bb();;
val it : mgoalstack =
`2 + 2 = x`
Metavariables: `x`,

# ee (CONCAT_SLV REFL_SLV (ACCEPT_SLV(ARITH_RULE `2 + 2 = 4`)));;
val it : mgoalstack = No sub(m)goals

# list_of_stream(top_thms());;
val it : (instantiation * thm) option list =
  [Some (([], [(`2 + 2`, `x`)], []), |- 2 + 2 = 2 + 2);
   Some (([], [(`4`, `x`)], []), |- 2 + 2 = 4)]
\end{verbatim}

The function
\verb|solve : solver -> term -> (term list * thm) stream|
runs the proof search non interactively and produces a list of
solutions as already shown in Section~\ref{sec:a-simple-example}.  In
this last case it would be
\begin{verbatim}
# list_of_stream (
    solve (CONCAT_SLV REFL_SLV (ACCEPT_SLV(ARITH_RULE `2 + 2 = 4`)))
          `??x. 2 + 2 = x`);;
val it : ((term * term) list * thm) list =
  [([(`2 + 2`, `x`)], |- 2 + 2 = 2 + 2); ([(`4`, `x`)], |- 2 + 2 = 4)]
\end{verbatim}



\section{Case study: Evaluation for a LISP-like language}
\label{sec:lisp-eval}

The material in this section is strongly inspired by the ingenious work of
Byrd, Holk, and Friedman about the miniKanren system
\cite{Byrd:2012:MLU:2661103.2661105}, where the authors work with the semantics
of the Scheme language.  Here we target a dynamically scoped variant of the
LISP language~--not unlike it is done in~\cite{10.1145/3110252}--~formalized as
an object language inside the HOL prover.  The HOL prover could be a powerful
tool for a formal study of the meta-theory of a programming language such as
LISP.  In this perspective, this section may have a scientific interest beyond
the entertaining nature of the example it is going to present.

First, we need to extend our HOL~Light environment with an object
datatype \verb|sexp| for encoding S-expressions according to the following BNF grammar
\begin{verbatim}
  sexp ::= Symbol string
         | List (sexp list)
\end{verbatim}
For instance, the sexp \verb|(list a (quote b))| is represented as HOL
term with
\begin{verbatim}
`List [Symbol "list";
       Symbol "a";
       List [Symbol "quote";
             Symbol "b"]]`
\end{verbatim}
This syntactic representation can be hard to read and gets quickly
cumbersome as the size of the terms grows.  Hence, we also introduce a
notation for concrete sexp terms, which is activated by the syntactic
pattern \verb|'(|\ldots\verb|)|.  For instance, the above example
is written in the HOL concrete syntax for terms as
\begin{verbatim}
`'(list a (quote b))`
\end{verbatim}

\noindent
With this setup, we can easily specify the evaluation rules for our
minimal LISP-like language.  This is a ternary
predicate~\verb|`|$\mathtt{EVAL}\ e\ x\ y\mathtt{}$\verb|`| which
satisfies the following clauses:
\begin{enumerate}
\item quoted expressions
\begin{verbatim}
# EVAL_QUOTED;;
|- !e q. EVAL e (List [Symbol "quote"; q]) q
\end{verbatim}
\item variables
\begin{verbatim}
# EVAL_SYMB;;
|- !e a x. RELASSOC a e x ==> EVAL e (Symbol a) x
\end{verbatim}
\item lambda abstractions
\begin{verbatim}
# EVAL_LAMBDA;;
|- !e l. EVAL e (List (CONS (Symbol "lambda") l))
                (List (CONS (Symbol "lambda") l))
\end{verbatim}
\item lists
\begin{verbatim}
# EVAL_LIST;;
|- !e l l'. ALL2 (EVAL e) l l'
            ==> EVAL e (List (CONS (Symbol "list") l)) (List l')
\end{verbatim}
\item unary applications
\begin{verbatim}
# EVAL_APP;;
|- !e f x x' v b y.
     EVAL e f (List [Symbol "lambda"; List[Symbol v]; b]) /\
     EVAL e x x' /\ EVAL (CONS (x',v) e) b y
     ==> EVAL e (List [f; x]) y
\end{verbatim}
\end{enumerate}
The predicate \verb|`EVAL`| is inductively definided, i.e., it is (informally)
the \emph{smallest} predicate that satisfies the above rules.

We now use our framework for running a certified evaluation process
for this language.  First, we define a solver for a single step of
computation
\begin{verbatim}
let STEP_SLV : solver =
  COLLECT_SLV
    [CONJ_SLV;
     ACCEPT_SLV EVAL_QUOTED;
     THEN_SLV (RULE_SLV EVAL_SYMB) RELASSOC_SLV;
     ACCEPT_SLV EVAL_LAMBDA;
     RULE_SLV EVAL_LIST;
     RULE_SLV EVAL_APP;
     ACCEPT_SLV ALL2_NIL;
     RULE_SLV ALL2_CONS];;
\end{verbatim}
In the above code, we collect the solutions of several different
solvers.  Other than the five rules of the \verb|EVAL| predicate, we
include specific solvers for conjunctions and the two predicates
\verb|REL_ASSOC| and \verb|ALL2|.

Let us mention that the definition of solvers such us \verb|STEP_SLV| above could be automatically derived from the set of clauses by performing a syntactical analysis.
However, we did not invest time so far on this kind of improvements,
since we are still experimenting with the basis of the system.

The top-level recursive solver for the whole evaluation predicate is now easy to define:
\begin{verbatim}
let rec EVAL_SLV : solver =
   fun g -> CONCAT_SLV ALL_SLV (THEN_SLV STEP_SLV EVAL_SLV) g;;
\end{verbatim}

Let us make a simple test.  The evaluation of the expression
\begin{verbatim}
((lambda (x) (list x x x)) (list))
\end{verbatim}
can be obtained as follows:
\begin{verbatim}
# get (solve EVAL_SLV
             `??ret. EVAL []
                          '((lambda (x) (list x x x)) (list))
                          ret`);;

val it : term list * thm =
  ([`ret = '(() () ())`],
   |- EVAL [] '((lambda (x) (list x x x)) (list)) '(() () ()))
\end{verbatim}

Again, we can use the declarative nature of logic programs to run the
computation backwards.  For instance, one intriguing exercise is the
generation of quine programs, that is, programs that evaluate to
themselves.  In our formalization, they are those terms $q$ satisfying
the relation \verb|`EVAL|~\verb|[]|~$q$~$q$\verb|`|.  The following command
computes the first two quines found by our solver.
\begin{verbatim}
# let sols = solve EVAL_SLV `??q. EVAL [] q q`);;
# take 2 sols;;

val it : (term list * thm) list =
  [([`q = List (Symbol "lambda" :: _3149670)`],
    |- EVAL [] (List (Symbol "lambda" :: _3149670))
       (List (Symbol "lambda" :: _3149670)));
   ([`q =
      List
      [List
       [Symbol "lambda"; List [Symbol _3220800];
        List [Symbol "list"; Symbol _3220800; Symbol _3220800]];
       List
       [Symbol "lambda"; List [Symbol _3220800];
        List [Symbol "list"; Symbol _3220800; Symbol _3220800]]]`],
    |- EVAL []
       (List
       [List
        [Symbol "lambda"; List [Symbol _3220800];
         List [Symbol "list"; Symbol _3220800; Symbol _3220800]];
        List
        [Symbol "lambda"; List [Symbol _3220800];
         List [Symbol "list"; Symbol _3220800; Symbol _3220800]]])
       (List
       [List
        [Symbol "lambda"; List [Symbol _3220800];
         List [Symbol "list"; Symbol _3220800; Symbol _3220800]];
        List
        [Symbol "lambda"; List [Symbol _3220800];
         List [Symbol "list"; Symbol _3220800; Symbol _3220800]]]))]
\end{verbatim}

One can easily observe that any lambda expression is trivially a quine
for our language.  This is indeed the first solution found by our
search:
\begin{verbatim}
([`q = List (Symbol "lambda" :: _3149670)`],
 |- EVAL []
         (List (Symbol "lambda" :: _3149670))
         (List (Symbol "lambda" :: _3149670)))
\end{verbatim}

The second solution is more interesting.  Unfortunately, it is
presented in a form that is hard to decipher.  A simple trick can help
us to present this term as a concrete sexp term: it is enough to
replace the HOL generated variable (\verb|`_3149670`|) with a concrete
string.  This can be done by an ad-hoc substitution:
\begin{verbatim}
# let [_; i2,s2] = take 2 sols;;
# vsubst [`"x"`,hd (frees (rand (hd i2)))] (hd i2);;

val it : term =
  `q = '((lambda (x) (list x x)) (lambda (x) (list x x)))`
\end{verbatim}

If we take one more solution from \verb|sols| stream, we get a new
quine which, interestingly enough, is precisely the one obtained in
\cite{Byrd:2012:MLU:2661103.2661105}:
\begin{verbatim}
val it : term =
  `q =
   '((quote (lambda (x) (list x (list (quote quote) x))))
     (quote (quote (lambda (x) (list x (list (quote quote) x))))))`
\end{verbatim}

\section{Description of our code}
\label{sec:code}

The HOL Light theorem prover and our extension are written in OCaml and, more
precisely, in a rather minimal and conservative subdialect of it, which should
be understandable to everyone that has some familiarity with any of the
languages of the ML family.  Our code is available from a public repository,
in particular, a release has been created at
\url{https://github.com/massimo-nocentini/kanren-light/releases/tag/miniKanren2020} .

Besides the code presented in this article, the above repository contains some
other experiments of various nature, including the following:
\begin{itemize}
  \item An implementation of the Quicksort algorithm.  The procedure outputs
	  the sorted list together with a formal proof that such list is indeed
		sorted and in bijection with the input lists.
  \item A solver for the \emph{Monte Carlo Lock}, a brain teaser by
	  Smullyan~\cite{smullyan2009lady}, where one has to unlock a
		\emph{safe} whose \emph{key} is the fixed point of an abstract
		machine.  The interesting thing is that the solver is
		essentially derived from the formal specification in HOL of the
		puzzle.
  \item An intuitionistic first-order tautology prover \verb|ITAUT_SLV|.  This
	  is inspired by a similar tactic \verb|ITAUT_TAC| already available in
		HOL Light.\footnote{The tactic \texttt{ITAUT\_TAC} has a
		peculiar role in the HOL system.  It is used during the
		\emph{bootstrap} of the system to prove several basic logical
		lemmas.  After the intial stages, a much more powerful and
		faster procedure for (classical) first-order logic
		\texttt{MESON\_TAC} is installed in the system, and the
		\texttt{ITAUT\_TAC} becomes superfluous.} However, HOL tactics
		cannot backtrack, which implies that \verb|ITAUT_TAC| is
		incomplete.  Our solver \verb|ITAUT_SLV| is coded in pretty
		much the same way as \verb|ITAUT_TAC|, but it is complete
		(although this latter fact can be claimed only via a
		meta-theoretical analysis).
\end{itemize}

With respect to the existing framework of HOL Light, our effort didn't apply
any change to both existing structures and computation flow, it just adds a parallel
way of proving things. The connection point is the \verb|type mgoal = term list * goal|
that enhances a \verb|goal| with a list of meta-variables and, eventually, all the
complexity of the presented work lies in their correct bookkeeping and in the handling
of goal streams.

Our code is conceived for experimenting, and very little
or no attention has been paid to optimizations.  Despite this, the OCaml
runtime and the HOL Light implementation have an established reputation of
being time- and memory-efficient systems (compared with similar tools).  From
our informal tests, it seems that this efficiency is, at least partially,
inherited by our implementation.

\section{Future and related work}
\label{sec:conclusions}

We presented a rudimentary framework inspired by miniKanren systems implemented on
top of the HOL~Light theorem prover that enables a logic programming paradigm
for proof searching.  More specifically, it facilitates the use of
meta-variables in HOL goals and permits backtracking during the proof
construction.
Despite the simplicity of the present implementation, we have already
shown the implementation of some paradigmatic examples of
logic-oriented proof strategies.

It would be interesting to enhance our framework with more features:
\begin{itemize}
\item Implement higher-order unification as Miller's higher-order patterns, so
	that our system can enable higher-order logic programming in the style
		of $\lambda$Prolog \cite{10.5555/648228.752141}.
\item Support constraint logic programming \cite{Hemann_2017}, e.g., by
	adapting the data structure that represents goals.
\end{itemize}

Besides extending our system with new features, we plan to test it on further
examples.  One natural domain of applications would be the development of
decision procedures.  While HOL Light already offers some remarkable tools for
automatic theorem proving, our system could offer new alternatives leaning to
simplicity and compositionality.  For instance, we could try to translate in
our system the approach
of~$\alpha$lean\textit{TAP}~\cite{10.1007/978-3-540-89982-2_26} for
implementing an automatic procedure for first-order classical logic in
HOL~Light analogous to the~\verb|blast| tactic of Paulson~\cite{Paulson1999} in
Isabelle.

\bibliographystyle{ACM-Reference-Format}
\bibliography{scilp}

\end{document}